\documentclass[prd,aps,superscriptaddress]{revtex4}
\usepackage{amssymb}
\usepackage{graphicx}
\usepackage{amsmath}

\begin{document}

\title{The photino sector and a confining potential in a
supersymetric Lorentz-symmetry-violating model}
\author{H. Belich}
\email{belichjr@gmail.com}
\affiliation{Departamento de F\'{\i}sica e Qu\'{\i}mica, Universidade Federal do Esp\'{\i}rito Santo, Vit\'{o}ria, ES, Brasil}
\author{L. D. Bernald}
\email{ldurand@cbpf.br}
\affiliation{Centro Brasileiro de Pesquisas F\'\i sicas, Rio de Janeiro, RJ, Brasil}
\author{Patricio Gaete}
\email{patricio.gaete@usm.cl}
\affiliation{Departmento de F\'{\i}sica and Centro Cient\'{\i}fico-Tecnol\'ogico de Valpara\'{\i}so, Universidad T\'{e}cnica Federico Santa Mar\'{\i}a, Valpara\'{\i}so, Chile}
\author{J. A. Helay\"el-Neto}
\email{helayel@cbpf.br}
\affiliation{Centro Brasileiro de Pesquisas F\'\i sicas, Rio de Janeiro, RJ, Brasil}
\date{\today }

\begin{abstract}
We study the spectrum of the minimal supersymmetric extension of the 
Carroll-Field-Jackiw model for Electrodynamics with a topological
Chern-Simons-like Lorentz-symmetry violating term. We identify a number of independent background fermion condensates, work out the gaugino dispersion relation and propose a photonic effective action to consider aspects of confinement induced by the SUSY background fermion condensates, 
which also appear to signal  Lorentz-symmetry violation in the photino sector of the action. Our calculations of the static potential are carried out within the framework of the gauge-invariant but path-dependent variables formalism which are alternative to the Wilson loop approach. Our results show that the interaction energy contains a linear term leading to the confinement of static probe charges. 
\end{abstract}

\maketitle

\section{Introduction}

The possibility that Lorentz and CPT symmetries be spontaneously broken at a very fundamental level, such as in the context of String Theory \cite{Colladay, Samuel}, has driven a very intensive activity and the so-called Standard Model Extension (SME) appears as a very suitable framework to probe details of 
Lorentz-symmetry violation (LSV) in diverse situations such as photon physics  \cite{photons1,photons2}, effects of radiative corrections  \cite{Radiative} , systems of fermions \cite{fermions}, neutrino physics   \cite{neutrinos}, topological defects \cite{Defects}, topological phases   \cite{Phases}, cosmic rays  \cite{CosmicRay}, particle decays \cite{Iltan} and a number of other relevant aspects of physical systems \cite{Lehnert1,General}. Also, different experiments have been proposed in connection with the SME that yield important stringent bounds on the parameters associated to systems where LSV is present  \cite{Tests, CPT}  .

The breaking of relativistic and CPT invariances have also been extensively studied in the framework of a modified Dirac theory  \cite{Hamilton} and its non-relativistic regime, with the calculation and discussion of the spectrum of the non-relativistic hydrogen atom \cite{Manojr}. In the direction of fermionic models in the presence of LSV, there has been an effort to associate magnetic properties of spinless and/or neutral particles if a non-minimal coupling of the Lorentz-symmetry violating background is taken into account. Still in the realm of atomic physics and optics, we should quote a line of works that set out to examine effects of LSV in electromagnetic cavities and optical systems \cite{Cavity, Masers} which have finally contributed to set up new bound on the parameters associated to LSV.

It is clear that the breaking of Lorentz symmetry takes place in the framework of a more fundamental physics, at a high energy scale. Whenever this fundamental physics is at work, SUSY might be exact or it could be broken in a scale close to the scale of this primary physics. Our assumption is that, in a high-energy regime, Lorentz symmetry violation (LSV) should not be disconnected from SUSY. We believe, and we adopt this viewpoint, that the scenario for the LSV is dominated by SUSY or, at least, is affected by imprints of an eventually broken SUSY. We cast, in the works of Ref. \cite{LSVSUSY}, a list of papers that introduce SUSY in direct association with models that take into account Lorentz- and CPT-symmetry violation. More recently, Redigolo \cite{David} proposes the criterium of renormalizability by weighted power-counting to build up superfield actions for Lorentz-violating supersymmetric quantum field theories.

So, in view of these considerations, SUSY is assumed to be present from the very outset of our proposal. This then means that LSV must be originated from some SUSY multiplet. This is the central idea of our whole approach. We shall see that this road takes us to a number of fermion condensates that characterize the background responsible for the LSV. In other words, going back to the supersymmetric regime, we claim it is possible to give a microscopic origin to the usually adopted vectors or tensors that parametrize the LSV background. They may show up as condensates of more fundamental fermions brought about by SUSY. Fermions associated to the latter condensate and directly induce new effects, such as mass splittings and extended dispersion relations for the photon and photino sectors.

Some technical aspects shall be described that yield the whole set of tensor backgrounds that are microscopically described in terms of fermions that condensate at the stage of LSV and, consequently, induce a SUSY breaking accompanied by the emergence of Goldstone-type fermions. No F- or D- type term is responsible for the SUSY spontaneous breaking in this case. LSV, whenever realized in a SUSY scenario, may naturally trigger SUSY breakdown.

We shall be giving the relevant details to understand the interplay between SUSY and LSV and we intend to be able to achieve, at the end, a way to introduce SUSY effects in extended photon and photino dispersion relations via the breaking of Lorentz symmetry. According to our proposal, SUSY effects come in through the fermions and other eventual bosons that accompany the Lorentz-symmetry violating background. Also, we may fix the scale of SUSY breaking by means of the well-known constraints that low-energy physics impose on the parameters that measure LSV. This is the content of the Sections II and III of our work.

On the other hand, one of the long-standing issues in gauge theories is a quantitative description of confinement. Meanwhile, phenomenological models still represent a key tool for understanding confinement physics, and can be considered as effective theories of QCD. In this last respect, we recall that many approaches to the problem of confinement rely on the phenomenon of condensation. For example, in the illustrative scenario of dual superconductivity, where it is conjectured that the QCD vacuum behaves as a dual type II superconductor. Thus, due to the condensation of magnetic monopoles, the chromo-electric field acting between $q\overline{q}$ pair is squeezed into strings, and the nonvanishing string tension represents the proportionality constant in the linear potential. With these considerations in mind, in previous works \cite{GaeteEuro3, GaeteHel}, 
we have studied, and we report on it in Section IV, the impact of condensates on physical observables in terms of the gauge-invariant but path-dependent variables formalism, both in $(3+1)$ and $(2+1)$ dimensions. In this perspective, the main goal of this paper is to consider the impact of fermionic condensates, which emerge after SUSY and Lorentz symmetries are broken in the supersymmetric Carroll-Field-Jackiw model \cite{CFJ}, on a physical observable.  As a result, we obtain that the potential energy is the sum of a Yukawa and a linear potential, leading to the confinement of static probe charges. Interestingly enough, the above static potential profile is analogous to that encountered in gluodynamics in curved spacetime \cite{GaeteEuro1}, as well as, for a non-Abelian gauge theory with a mixture of pseudoscalar and scalar couplings \cite{GaeteEuro2}. The above result reveals a new equivalence between effective gauge theories, in spite of the fact that they have different constraint structures. Our concluding remarks are presented in Section V.

\section{The Supersymmetric Extension of the Maxwell-Chern-Simons Model}

We begin with the modified supersymmetric Abelian gauge model proposed in \cite{Baeta}, which is a superfield version of the Carroll-Field-Jackiw Electrodynamics \cite{CFJ} with a background superfield that realizes the Lorentz-symmetry breaking.
This model has the interesting property of preserving supersymmetry at the action level, while the Lorentz symmetry is violated in the sense of particle transformations. Adopting a covariant superspace-superfield
formulation, as it is given in \cite{Baeta}, we write:
\begin{equation}
S=\int d^{4}xd^{2}\theta d^{2}\bar{\theta}\left\{ W^{a}(D_{a}V)S+\overline{W}
_{\dot{a}}(\overline{D}^{\dot{a}}V)\overline{S}\right\} ,  \label{superjac}
\end{equation}
where the superfields $W_{a}$, $V$, $S$ and the susy-covariant derivatives, $
\ D_{a}$, $\overline{D}_{\dot{a}}$, are expressed as it follows below:
\begin{eqnarray}
D_{a} &=&\frac{\partial }{\partial \theta ^{a}}+i{\sigma ^{\mu }}_{a\dot{a}}
\bar{\theta}^{\dot{a}}\partial _{\mu } \\
\overline{D}_{\dot{a}} &=&-\frac{\partial }{\partial \bar{\theta}^{\dot{a}}}
-i\theta ^{a}{\sigma ^{\mu }}_{a\dot{a}}\partial _{\mu };
\end{eqnarray}
the field-strength superfield, $W_a$, is given by
\begin{equation}
W_{a}(x,\theta ,\bar{\theta})=-\frac{1}{4}\overline{D}^{2}D_{a}V.
\end{equation}
The action of eq. (\ref{superjac}) is gauge invariant up to surface terms. The Bianchi identities $D^a W_a  = \bar D_{\dot a} \bar W^{\dot a}  = 0$ and the constraints on $S$ and $\bar S$ (given in the sequel) ensure gauge invariance of our Lorentz-symmetry violating action. $W_a$ can be
$\theta $-expanded as below: 
\begin{eqnarray}
W_{a}(x,\theta ,\bar{\theta}) &=&\lambda _{a}\left( x\right) +i\theta ^{b}{\
\sigma ^{\mu }}_{b\dot{a}}\bar{\theta}^{\dot{a}}\partial _{\mu }\lambda
_{a}\left( x\right) -\frac{1}{4}{\bar{\theta}}^{2}\theta ^{2}\square \lambda
_{a}\left( x\right) +2\theta _{a}D\left( x\right) -i\theta ^{2}\bar{\theta}^{
\dot{a}}{\sigma ^{\mu }}_{a\dot{a}}\partial _{\mu }D\left( x\right) +{{{
\sigma }^{\mu \nu }}_{a}}^{b}\theta _{b}F_{\mu \nu }\left( x\right)   \notag
\\
&-&\frac{i}{2}{{{\sigma }^{\mu \nu }}_{a}}^{b}{\sigma ^{\alpha }}_{b\dot{a}
}\theta ^{2}\overline{\theta }^{\dot{a}}\partial _{\alpha }F_{\mu \nu
}\left( x\right) -i\sigma _{a\dot{a}}^{\mu }\partial _{\mu }\text{ }
\overline{\lambda }^{\dot{a}}\left( x\right) \theta ^{2},
\end{eqnarray}
and $V=V^{\dagger }$ is the so-called gauge-potential superfield, which is a real scalar. The Wess-Zumino gauge choice is adopted, as usually done if we are to perform component-field calculations: 
\begin{equation}
\text{ }V_{WZ}=\theta \sigma ^{\mu }\bar{\theta}A_{\mu }(x)+\theta ^{2}\bar{
\theta}\overline{\lambda }\left( x\right) +\bar{\theta}^{2}\theta \lambda
(x)+\theta ^{2}\bar{\theta}^{2}D.
\end{equation}
The background superfield, $S$, is so chosen to be a chiral supermultiplet. Such a constraint restricts the
highest spin component of the background to be an $s$ $=$ $\frac{1}{2}$
component-field. Also, according to the action of eq.(\ref{superjac}), one should notice that $S$ happens to be dimensionless. As a physical propagating superfield, its mass dimension would be equal to $1$. The $\theta$-expansion for the background superfield $S$ then reads:  
\begin{equation}
\overline{D}_{\dot{a}}S\left( x\right) =0,
\end{equation}
and 
\begin{equation}
S\left( x\right) \text{ }=s\left( x\right) +i\theta \sigma ^{\mu }\overline{
\theta }\partial _{\mu }s\left( x\right) -\frac{1}{4}{\bar{\theta}}
^{2}\theta ^{2}\square s\left( x\right) +\sqrt{2}\theta \psi \left( x\right)
+\frac{i}{\sqrt{2}}\theta ^{2}\overline{\theta }\overline{\sigma }_{\mu
}\partial _{\mu }\psi \left( x\right) +\theta ^{2}F\left( x\right) .
\end{equation}
The component-wise counterpart for the action (\ref{superjac}) is given by the expression (after the supersymmetric Maxwell action is also included): 
\begin{eqnarray}
S_{comp.} &=&\int {d^{4}x}\Biggl[-\left\{ {\frac{1}{4}+\frac{{\left( {
s+s^{\ast }}\right) }}{2}}\right\} F_{\mu \nu }F^{\mu \nu }+\frac{i}{2}
\partial _{\mu }\left( {s-s^{\ast }}\right) \varepsilon ^{\mu \alpha \beta
\nu }F_{\alpha \beta }A_{\nu }+\left\{ {\frac{1}{2}+4\left( {s+s^{\ast }}
\right) }\right\} D^{2}  \notag \\
&-&\left( {\frac{1}{2}-2s}\right) i\lambda \sigma ^{\mu }\partial _{\mu }
\bar{\lambda}-\left( {\frac{1}{2}-2s^{\ast }}\right) i\bar{\lambda}\bar{
\sigma}^{\mu }\partial _{\mu }\lambda -\sqrt{2}\lambda \sigma ^{\mu \nu
}\psi F_{\mu \nu }+\sqrt{2}\bar{\lambda}\bar{\sigma}^{\mu \nu }\bar{\psi}
F_{\mu \nu }  \notag \\
&+&\lambda \lambda F+\bar{\lambda}\bar{\lambda}F^{\ast }-2\sqrt{2}\lambda
\psi D-2\sqrt{2}\bar{\lambda}\bar{\psi}D\Biggr]\ .
\end{eqnarray}
By suitably choosing $s$, such that $s+s^{\ast }=0$ , it is the imaginary part of $s$ the
responsible for the appearance of the vector $v_{\mu }$ of the
Carroll-Field-Jackiw term; $s-s^{\ast }=-(i/2)v_{\mu }x^{\mu }$. D is fixed by its algebraic field
equation, $D=\sqrt{2}(\lambda \psi +\bar{\lambda}\bar{\psi})$. Making
use of Fierz rearrangements in all the $4$-fermion terms and rewriting the action in terms of Majorana 4-component spinors, we arrive at 
\begin{eqnarray}
S_{comp}=\int d^{4}x &&\left[ -\frac{1}{4}F_{\mu \nu }F^{\mu \nu }+\frac{1}{4}
v_{\mu }\varepsilon ^{\mu \alpha \beta \nu }F_{\alpha \beta }A_{\nu }-\frac{
i }{2}\bar{\Lambda}\bar{\gamma}^{\mu }\partial _{\mu }\Lambda \right. \notag \left. +\left( \text{
Re} (F)+\frac{1}{4}\bar{\Psi}\Psi \right) \bar{\Lambda}\Lambda -i\left( \text{Im}
(F)+\frac{1}{4}i \bar{\Psi}\gamma _{5}\Psi \right) \bar{\Lambda}\gamma _{5}\Lambda
\right.  \notag \\
&&-\frac{1}{4}\left(v_\mu+ \bar{\Psi}\gamma _{\mu }\gamma _{5}\Psi \right) (\bar{\Lambda}
\gamma ^{\mu }\gamma _{5}\Lambda )+\left. \sqrt{2}\bar{\Lambda}\Sigma ^{\mu
\nu }\gamma _{5}\Psi F_{\mu \nu }\right], 
\label{ecu}
\end{eqnarray}
with the Majorana fermions, $\Lambda$ (the gaugino) and $\Psi$ (the background fermion), given by:
\begin{equation}
\Lambda \equiv \left( 
\begin{array}{c}
\lambda _{\alpha } \\ 
\bar{\lambda}_{\dot{\alpha}}
\end{array}
\right) ,\Psi \equiv \left( 
\begin{array}{c}
\psi _{\alpha } \\ 
\bar{\psi}_{\dot{\alpha}}
\end{array}
\right), 
\end{equation}
and
\begin{equation}
\Sigma ^{\mu \nu }  \equiv \frac{i}{4}\left[ {\gamma ^\mu  ,\gamma ^\nu  } \right].
\end{equation}
It becomes clear, in eq. (\ref{ecu}), how the fermionic background, $\Psi$, and the scalar, $F$, affect the photino sector of SUSY in eq. (\ref{ecu}): they yield mass-type terms for $\Lambda$.
Also, we highlight the presence of a new photon-photino term (the very last term of (\ref{ecu})), which appears due exclusively to the presence of the fermionic component of the background.

By inspecting the SUSY transformations of the component fields, we notice that the breaking of
Lorentz symmetry necessarily implies the appearance of a sort of Goldstino particle,
since $\partial _{\mu }B$, being non-trivial, yields $\delta \Psi \neq 0$, once the SUSY variation of $\Psi$ reads as follows:
\begin{equation}
\delta \Psi =\partial _{\mu }(A-\gamma _{5}B)\gamma ^{\mu }\varepsilon
+f\varepsilon +g\gamma _{5}\varepsilon ,
\end{equation}
where $A$, $B$, $f$ and $g$ are such that $s=A+iB$, $F=f+ig$. $\varepsilon $ is the four-component
Majorana parameter of the SUSY transformation. This signals the presence of a Goldstone fermion produced as a perturbation around the background, even if $f=g=0$, but with $\partial _\mu  B =  - \frac{1}{4}v_\mu$. So, SUSY is broken together with Lorentz symmetry. Translations are not broken, since $v^{\mu}$ is constant and then no explicit $x^{\mu}$-dependence is present in (\ref{ecu}) through the background components fields ($
\Psi$ and $F$ are also  $x^{\mu}$-independent). So, Poincar\'e symmetry is actually broken in the sector of boosts and space rotations.

\section{The Photon and Photino Dispersion Relations and An Effective Photonic Action}

If we wish to read off the photon and photino dispersion relations in the presence of the complete background responsible for the LSV, namely $ 
\left\{ {A,B,\Psi _\alpha  ,f,g} \right\}$, it is suitable to express the the kinetic part of the Lagrangian, taking account that the background fields are fixed, in the form that is cast below:
\begin{equation}
{\mathcal{L}}=\frac{1}{2}{\Phi ^{t}\mathcal{O}}\Phi =\frac{1}{2}\left( 
\begin{array}{cc}
\bar{\Lambda}_{a} & A_{\mu }
\end{array}
\right) \left( 
\begin{array}{cc}
J^{ab} & L^{a\nu } \\ 
M^{\mu b} & N^{\mu \nu }
\end{array}
\right) \left( 
\begin{array}{c}
\Lambda _{b} \\ 
A_{\nu }
\end{array}
\right) ,  \label{eq:operator}
\end{equation}
where 
\begin{equation}
J^{ab} =-i(\gamma ^{\mu }\partial _{\mu })^{ab}+(2\text{Re}(F)+\frac{\mu}{2})1_{4\times 4}-i(2\text{Im}(F)+i\frac{\tau}{2})\gamma
_{5}^{ab}-\frac{1}{2}\left( v_\mu+\bar{\Psi}\gamma _{\mu }\gamma _{5}\Psi \right) \left(
\gamma ^{\mu }\gamma _{5}\right) ^{ab}, \label{vina1}
\end{equation}
\begin{equation}
L^{a\nu }  = 2\sqrt 2 \left( {\Sigma ^{\mu \nu } \gamma _5 } \right)^{ab} \Psi \partial _\mu, \label{vina2}
\end{equation}
\begin{equation}
M^{\mu b} =2\sqrt{2}\bar{\Psi}(\Sigma ^{\nu \mu }\gamma _{5})^{ab}\partial_{\nu }, \label{vina3}
\end{equation}
and
\begin{equation}
N^{\mu \nu } =\Box \theta ^{\mu \nu }-v_{\rho }\varepsilon ^{\rho \lambda
\mu \nu }\partial _{\lambda }-\frac{1}{\alpha }\Box \omega ^{\mu \nu }. \label{vina4}
\end{equation}
As we usually proceed, a gauge-fixing term with parameter $\alpha $ is introduced to ensure invertibility of $N$. In case we wished to explicitly read off the photon-photino propagators, we would have to compute ${\cal O}^ {-1}$. Since $J$ is invertible, ${\cal O}$ becomes non-singular whenever $N$ is also invertible. In (\ref{vina1})-(\ref{vina4}), we have defined 3 background fermion condensates:
\begin{equation}
\mu  \equiv \bar \Psi \Psi, \label{valpo-a}
\end{equation}
\begin{equation}
\tau  \equiv \bar \Psi \gamma _5 \Psi, \label{valpo-b}
\end{equation}
\begin{equation}
C^\mu   \equiv \bar \Psi \gamma ^\mu  \gamma _5 \Psi. \label{valpo-c}
\end{equation}

Since $\Psi $ is a Majorana spinor, we can ensure that $\mu $ is real, $\tau $ is purely imaginary and $B^{\mu }$ is a pseudo-vector with real components.
Upon some Fierzings and by considering that the Majorana $\Psi $-components are Grassmann-valued, we can readily show that 
\begin{equation}
\mu ^{2}=-\tau ^{2}=\frac{1}{4}C_{\mu }C^{\mu }.  \label{III-60}
\end{equation}
These relations have some important consequences:

\begin{itemize}
\item $C_{\mu}$ cannot be space-like, once $\mu ^2  =  - \tau ^2  \ge 0$;

\item $C_{\mu}=0$ yields $\mu=\tau=0$, so that no condensates would survive;

\item if $\mu=\tau=0$, then $C_{\mu}$ is light-like;

\item $C_{\mu}$ time-like implies $\mu\neq0$ and $\tau\neq0$. In this case, all condensates simultaneously contribute.
\end{itemize}

From (\ref{eq:operator}), we are ready to write down the photino dispersion relations,
\begin{equation}
\det \left( {J - LN^{ - 1} M} \right) = 0, \label{disp1}
\end{equation}
and the corresponding photon dispersion relations,
\begin{equation}
\det \left( {N - MJ^{ - 1} L} \right) = 0. \label{disp2}
\end{equation}

The fermionic opertor $J$ in (\ref{vina1}) is invertible. We compute $J^{-1}$ and quote its expression as follows:
\begin{equation}
J^{ - 1}  = A1_{4 \times 4}  + B\gamma _5  + R_\mu  \gamma ^\mu   + S_\mu  \gamma ^\mu  \gamma _5  + L_{\mu \nu } \Sigma ^{\mu \nu }, \label{disp3}
\end{equation}
with the coefficients $A, B, R_{\mu}, S_{\mu}$ and $L_{\mu \nu }=-L_{\nu \mu }$ listed below:

\begin{equation}
A=\left( {2{\mathop{\rm Re}\nolimits}\left( F\right) +\frac{\mu }{2}}\right)
\left( {4\left\vert F\right\vert ^{2}+\frac{3}{2}\mu ^{2}+2\delta -p^{2}+
\frac{{v^{2}}}{4}+\frac{1}{2}\left( {v\cdot C}\right) }\right) /\Delta ,
\label{III-75a}
\end{equation}
\begin{equation}
B=i\left( {2{\mathop{\rm Im}\nolimits}\left( F\right) +i\frac{\mu }{2}}
\right) \left( {4\left\vert F\right\vert ^{2}+\frac{3}{2}\mu ^{2}+2\delta
-p^{2}+\frac{{v^{2}}}{4}+\frac{1}{2}\left( {v\cdot C}\right) }\right)
/\Delta ,  \label{III-75b}
\end{equation}
\begin{equation}
R_{\mu }=\left[ {\left( {\frac{{p^{2}}}{2}+\frac{{v^{2}}}{8}+\frac{{\mu ^{2}}
}{4}+\frac{{\left( {v\cdot C}\right) }}{4}-2\left\vert F\right\vert
^{2}-\delta }\right) 2p_{\mu }-\frac{{\left\{ {\left( {p\cdot v}\right)
+\left( {p\cdot C}\right) }\right\} }}{2}\left( {v_{\mu }+C_{\mu }}\right) }
\right] /\Delta ,  \label{III-75c}
\end{equation}
\begin{equation}
S_{\mu }=\left[ {\left( {\frac{{p^{2}}}{2}+\frac{{v^{2}}}{8}+\frac{3{\mu ^{2}}
}{4}+\frac{{\left( {v\cdot C}\right) }}{4}+2\left\vert F\right\vert
^{2}+\delta }\right) \left( {v_{\mu }+C_{\mu }}\right) -\left\{ {\left( {
p\cdot v}\right) +\left( {p\cdot C}\right) }\right\} p_{\mu }}\right]
/\Delta ,  \label{III-75d}
\end{equation}
\begin{equation}
L_{\mu \nu }=\left[ {-2\left( {2{\mathop{\rm Im}\nolimits}\left( F\right) +
\frac{i}{2}\tau }\right) \left( {p_{\mu }v_{\nu }+p_{\mu }C_{\nu }}\right)
+\left( {2Re\left( F\right) +\frac{1}{2}\mu }\right) \left( {p_{\alpha
}v_{\beta }+p_{\alpha }C_{\beta }}\right) 
\varepsilon _{\mu \nu } \,^{\alpha \beta }}\right] /\Delta ,  \label{III-75e}
\end{equation}
\begin{equation}
\Delta =p^{4}-p^{2}\left[ {8\left\vert F\right\vert ^{2}+4\delta -\frac{{
v^{2}}}{2}-\left( {v\cdot C}\right) }\right] -2\left( {p\cdot v}\right)
\left( {p\cdot C}\right) -\left( {p\cdot v}\right) ^{2}+\left[ {4\left\vert
F\right\vert ^{2}+\frac{3}{2}\mu ^{2}+2\delta +\frac{{v^{2}}}{4}+\frac{{
\left( {v\cdot C}\right) }}{2}}\right] ^{2},  \label{III-75f}
\end{equation}
\begin{equation}
where \, \delta \equiv{\mathop{\rm Re}\nolimits}\left( F\right) \mu +i{\mathop{\rm Im}
\nolimits}\left( F\right) \tau.  \label{III-75g}
\end{equation}

Now, that we know $J^{-1}$, we can rewrite the photino dispersion relation (\ref{disp1}) according to
\begin{equation}
\det \left( {J - LN^{ - 1} M} \right) = \left( {\det J} \right)\left[ {\det \left( {1 - J^{ - 1} LN^{ - 1} M} \right)} \right] = 0. \label{disp3}
\end{equation}
Since $(1 - J^{ - 1} LN^{ - 1} M)$ is invertible, the photino dispersion relation reduces to
\begin{equation}
\det J = \Delta  = 0, \label{disp4}
\end{equation}
with $\Delta$ given by (\ref{III-75f}).
This expression then brings to light how the background vector and scalar, $v^{\mu}$ and $F$, and the fermion condensates, $\mu$, $\tau$ and $B^{\mu}$, combine to govern the photino propagation modes.

So long as the photon is concerned, its dispersion relation (\ref{disp2}) can be re-organized as ($N$ is invertible):
\begin{equation}
\det \left( {N - MJ^{ - 1} L} \right) = \left( {\det N} \right)\left[ {\det \left( {1 - N^{ - 1} MJ^{ - 1} L} \right)} \right] = 0. \label{disp5}
\end{equation}
Again, $({1 - N^{ - 1} MJ^{ - 1} L})$ is non-singular, so that
\begin{equation}
det N=0 \label{disp6}
\end{equation}
responds for the photon dispersion relation \cite{CFJ}:
\begin{equation}
p^4  + v^2 p^2  - \left( {v \cdot p} \right)^2  = 0. \label{disp7}
\end{equation}
So, only the background vector $v^{\mu}$ actually affects the photon propagating modes. The scalar background, $F$, and the fermion condensates $\mu$, $\tau$, $B^{\mu}$, do not change the photon propagating modes of the non-supersymmetric Carroll-Field-Jackiw model. However, it is worthy mentioning that, even if the mixing operators, ${L}$ and $M$, in ${\cal O}$ do not contribute to both the photon and photino dispersion relations, we point out that they do affect the propagators of the photon-photino sector and they are very relevant for the analysis of the residue matrices of the $ 
\left\langle {\bar \Lambda _\alpha  \Lambda _\beta  } \right\rangle$-, $ 
\left\langle {\bar \Lambda _\alpha  A_\mu  } \right\rangle$- and $ 
\left\langle {A_\mu  A_\nu  } \right\rangle$-propagators at their poles. The latter are clearly the zeroes of the equations that give the dispersion relations, and this becomes clear since the propagators above can be read off from the matrix ${\cal O}^{-1}$ ( ${\cal O}$ given in eq.(\ref{eq:operator})), whose general form can be organized as follows:
\begin{equation}
{\cal O}^{ - 1}  = \left( {\begin{array}{*{20}c}
   X & Y  \\
   Z & W  \\
\end{array}} \right), \label{disp8}
\end{equation}
where
\begin{equation}
X \equiv \left( {J - LN^{ - 1} M} \right)^{ - 1}  = \left( {1 - J^{ - 1} LN^{ - 1} M} \right)^{ - 1} J^{ - 1}, \label{disp9}
\end{equation}
\begin{equation}
W \equiv \left( {N - MJ^{ - 1} L} \right)^{ - 1}  = \left( {1 - N^{ - 1} MJ^{ - 1} L} \right)^{ - 1} N^{ - 1} . \label{disp10}
\end{equation}
\begin{equation}
Y \equiv  - J^{ - 1} L\left( {N - MJ^{ - 1} L} \right)^{ - 1}  =  - J^{ - 1} L\left( {1 - N^{ - 1} MJ^{ - 1} L} \right)N^{ - 1}, \label{disp11}
\end{equation}
\begin{equation}
Z \equiv  - N^{ - 1} M\left( {J - LN^{ - 1} M} \right)^{ - 1}  =  - N^{ - 1} M\left( {1 - J^{ - 1} LN^{ - 1} M} \right)^{ - 1} J^{ - 1} . \label{disp12} 
\end{equation}
Eqs.(\ref{disp9}) and (\ref{disp10}) clearly confirm that the propagator poles, that are accomodated in $J^{-1}$ and $N^{-1}$, exactly correspond to the zeroes of dispersion relations (\ref{III-75f}) and (\ref{disp7}). For the sake of our discussions in this work, we do not need to explicitly compute the propagators. We are only interested in working out the dispersion relations; this is why we do not carry out the explicit calculation of ${\cal O}^{-1}$.

From (\ref{III-75f}) and (\ref{disp7}), we see that only $v^{\mu}$ enters the photon dispersion relation though it also enters the photino dispersion relation. So, let us consider the particular situation
\begin{equation}
\Psi  = 0, \label{pgd1}
\end{equation} 
and
\begin{equation}
F=0, \label{pgd1A}
\end{equation}
so that all fermion condensates are switched off. In such a case, the photino dispersion relation simplifies to
\begin{equation}
\Delta  = p^4  - \frac{1}{2}v^2 p^2  - \left( {v \cdot p} \right)^2  + \frac{1}{16}v^4. \label{pgd2}
\end{equation}
It then becomes clear that a massless photon (according to (\ref{disp7}), characterized by $ {v \cdot p}=0$) is not  accompanied by a massless photino, since $ {v \cdot p}=0$ is not a zero of $\Delta$ whenever $p^{2}=0$. This confirms that the LSV actually induces a SUSY breaking, by splitting the photon and photino masses. In the special case of a space-like $v^\mu$ (see Klinkhamer in Ref. \cite{photons1}), a massless photon is accompanied by a massive photino whose mass is calculated to be
\begin{equation}
m_{photino}  = \frac{1}{{\sqrt 2 }} \left| {\vec v} \right|, \label{phothin}
\end{equation}
where $\vec v$ is the spatial component of $v^\mu$. In this particular situation, the photon-photino mass splitting is directly measured by $v^\mu$.

On the other hand, if $v^{\mu}=0$ and the fermions condensates are non-trivial, $p^{2}=0$ is always a zero of (\ref{disp7}) (so, a massless photon is present in the spectrum in such a case), but it is never a zero of $\Delta$, so that, in this special case, a massless photino never shows up, which is again compatible with the situation of broken SUSY.

Finally, we can work to get a photonic effective action by integrating out the photino field. To do that, we are allowed to redefine $\Lambda$ according to the shift:
\begin{equation}
\Upsilon=\Lambda +{{J}}^{-1}\sqrt{2}\Sigma^{\mu\nu}\gamma_5 \Psi F_{\mu\nu},  \label{disp13}
\end{equation}
and 
\begin{equation}
\bar{\Upsilon}=\bar{\Lambda} -\sqrt{2}\bar{\Psi}\gamma_5\Sigma^{\mu\nu}\bar{{{J}}}^{-1}F_{\mu\nu},  \label{disp14}
\end{equation}
where $J^{-1}$  and ${\bar J^{-1}}$ already explicitly computed. Though there is a manifest non-locality in the field reshufflings (\ref{disp13}) and (\ref{disp14}),
this is harmless so long as we are interested in reading off an effective action for the photon by eliminating the $\Lambda _\alpha$-$A^{\mu}$ mixing and integrating out the fermions in the action (\ref{ecu}).

With the explicit expressions for $J^{-1}$ and ${\bar J^{-1}}$ and by means of manipulations with the $\gamma ^{\mu }$-algebra, we are able to cast the form of the photon effective action as given below: 
\begin{eqnarray}
{\cal L} &=&  - \frac{1}{4}F_{\mu \nu }^2  + \frac{1}{4}\varepsilon ^{\mu \nu \alpha \beta } v_\mu  A_\nu  F_{\alpha \beta }  + F_{\mu \nu } \left( {\frac{1}{4}\mu A + \frac{1}{4}\tau B + \frac{1}{2}C_\rho  S^\rho  } \right)F^{\mu \nu }  + F_{\mu \lambda } \left( {2C^\mu  S_\nu} \right)F^{\lambda \nu }  \nonumber\\
&+& F_{\mu \nu } \left( {\frac{i}{2}\tau A + \frac{i}{2}\mu B} \right)\tilde F^{\mu \nu }  - i\tilde F_{\mu \rho } C^\rho  R_\sigma  F^{\mu \sigma }  - iF_{\mu \rho } C^\rho  R_\sigma  \tilde F^{\mu \sigma }  - \frac{1}{2}\tau F_{\mu \lambda } L_\nu  ^\lambda  \tilde F^{\mu \nu }.  \label{III-80}
\end{eqnarray}

It is remarkable to point out that the breaking of Lorentz
symmetry naturally induces axionic-like terms, $F\tilde{F}$, whose coefficients are originated from background fermion condensates. In (\ref{III-80}), we warn that, in all the coefficients, $A$, $B$, $R_{\mu}$, $S_{\mu}$ and $L_{\mu\nu}$, the terms where there appear a $4$-momentum, $p^{\mu}$ and $p^2$ are to be understood as written down in coordinate space ($p_{\mu}=i\partial _\mu$).

\section{Interaction energy}

We shall now calculate the static potential using the gauge-invariant but path-dependent variables formalism along the lines of Refs.\cite{GaeteSch, GaeteEuro1, GaeteEuro2, Gaete99}. To this end, we will compute the expectation value of the energy operator $H$ in the physical state $|\Phi\rangle$ describing the sources, which we will denote by ${\langle H \rangle}_{\Phi}$. To carry out our study, we consider the effective Lagrangian density for $A_\mu$ which is given by Eq. (\ref{III-80}). Next, by considering the special case $v^\mu=0$, $F\neq0$, $\delta\neq0$, $B_i=0$
and $B_0\neq0$, the effective Lagrangian becomes 
\begin{equation}
\mathcal{L} = - \frac{1}{4}F_{\mu \nu } \left[ {\frac{{\nabla ^4 - a^2
\nabla ^2 + b^2 }}{{\nabla ^4 - m_1 \nabla ^2 + m_2^2 }}} \right]F^{\mu \nu } + \frac{{C_0^2 }}{2} F_{0i} \left[ {\frac{{\nabla ^2 + m_2 }}{{\nabla ^4 + m_1
\nabla ^2 + m_2^2 }}} \right]F^{0i} + \frac{Q}{2}F_{\mu \nu } \left[ {\frac{{
 {\nabla ^2 - m_2 } }} {{\nabla ^4 + m_1 \nabla ^2 + m_2^2 }}} \right]
\tilde F^{\mu \nu },  \label{WLIV05}
\end{equation}
Here, $a^2 \equiv m_1 -
\left( {2P + C_0^2 } \right)$, $b^2 \equiv m_2 \left( {m_2 - C_0^2 }
\right) + 2 m_2P$, $m_1 \equiv 8\left| F \right|^2 + 4\delta$, $m_2
\equiv 4\left| F \right|^2 + \frac{3}{2}\mu ^2 + 2\delta$, $P \equiv \mu
\left( {2{\mathop{\rm
Re}\nolimits} \left( F \right) + \frac{\mu }{2}} \right) + i\left( {2{
\mathop{\rm Im}\nolimits} \left( F \right) + i\frac{\tau }{2}} \right)\tau$
and $Q \equiv 2i{\mathop{\rm Re} \nolimits} \left( F \right)\tau - 2{
\mathop{\rm Im}\nolimits} \left( F \right)\mu$. Before going into details, we recall that this paper is aimed at studying the static potential. In such a case, we can drop out terms with time derivatives in the system described by Eq.(\ref{WLIV05}). With this remark, the canonical quantization of this theory, from the Hamiltonian point of view, readily follows. The Hamiltonian analysis starts with the computation of the canonical momenta, $\Pi^0=0$ and $
\Pi ^i = - \frac{{\left( {\nabla ^4 - \xi ^2 \nabla ^2 + \rho ^2 } \right)}}{{
\left( {\nabla ^4 - m_1 \nabla ^2 + m_2^2 } \right)}}F^{0i} - Q\frac{{\left( {
\nabla ^2 - m_2 } \right)}}{{\left( {\nabla ^4 - m_1 \nabla ^2 + m_2^2 } \right)}}B^i$, where $\xi ^2 \equiv a^2 +C_0^2$, $\rho ^2 \equiv b^2 - m_2C_0^2$ and $B^k = - {\textstyle{\frac{1 }{2}}}
\varepsilon ^{kij} F_{ij}$. The canonical Hamiltonian corresponding to (\ref
{WLIV05}) is 
\begin{eqnarray}
H_C &=& \int {d^3 x}\Biggl[- A_0 \partial _i \Pi ^i - \frac{1}{2}\Pi _i
\left( {\frac{{\nabla ^4 - m_1 \nabla ^2 + m_2^2 }}{{\nabla ^4 - \xi ^2 \nabla ^2 + \rho ^2 }}} \right)\Pi ^i -\frac{Q}{2}\Pi _i \left( {\frac{{\nabla ^2 - m_2 }}{{\nabla ^4 - \xi ^2 \nabla ^2 + \rho ^2 }}} \right)B^i + QB_i \Pi ^i  \notag \\
&+& Q^2 B_i \frac{{\left( {\nabla ^2 - m_2 } \right)}}{{\left( {\nabla ^4 -
m_1 \nabla ^2 + m_2^2 } \right)}}B^i + \frac{1}{4}F_{ij}
\left( {\frac{{\nabla ^4 - a^2 \nabla ^2 + b^2 }}{{\nabla ^4 - m_1 \nabla ^2 +
m_2 ^2 }}} \right)F^{ij}\Biggr]\ .  \label{WLIV10}
\end{eqnarray}

Demanding that the primary constraint $\Pi_0=0$ be preserved in the course
of time, one obtains the secondary Gauss law constraint of the theory as $
\Gamma_1\equiv \partial _i \Pi ^i =0$. The preservation of $\Gamma_1$ for
all times does not give rise to any further constraints. The theory is thus
seen to possess only two constraints, which are first class, therefore the
theory described by $(\ref{WLIV05})$ is a gauge-invariant one. The extended
Hamiltonian that generates translations in time then reads $H = H_C + \int {d^3 } x\left( {c_0 \left( x \right)\Pi _0 \left( x \right) + c_1 \left( x
\right)\Gamma _1 \left( x \right)} \right)$, where $c_0 \left( x \right)$
and $c_1 \left( x \right)$ are the Lagrange multiplier fields. Moreover, it
is straightforward to see that $\dot{A}_0 \left( x \right)= \left[ {A_0
\left( x \right),H} \right] = c_0 \left( x \right)$, which is an arbitrary
function. Since $\Pi^0 = 0$ always, neither $A^0 $ nor $\Pi^0 $ are of
interest in describing the system and may be discarded from the theory. Thus
the Hamiltonian takes the form 
\begin{eqnarray}
H &=& \int {d^3 x}\Biggl[c(x) \left( {\partial _i \Pi ^i} \right) -  \frac{1}{2}\Pi _i
\left( {\frac{{\nabla ^4 - m_1 \nabla ^2 + m_2^2 }}{{\nabla ^4 - \xi ^2 \nabla ^2 + \rho ^2 }}} \right)\Pi ^i -\frac{Q}{2}\Pi _i \left( {\frac{{\nabla ^2 - m_2 }}{{\nabla ^4 - \xi ^2 \nabla ^2 + \rho ^2 }}} \right)B^i + QB_i \Pi ^i  \notag \\
&+& Q^2 B_i \frac{{\left( {\nabla ^2 - m_2 } \right)}}{{\left( {\nabla ^4 -
m_1 \nabla ^2 + m_2^2 } \right)}}B^i + \frac{1}{4}F_{ij}
\left( {\frac{{\nabla ^4 - a^2 \nabla ^2 + b^2 }}{{\nabla ^4 - m_1 \nabla ^2 +
m_2 ^2 }}} \right)F^{ij}\Biggr]\ ,  \label{WLIV15}
\end{eqnarray}
where $c(x) = c_1 (x) - A_0 (x)$.

The quantization of the theory requires the removal of non-physical
variables, which is accomplished by imposing a gauge condition such that the
full set of constraints becomes second class. A particularly convenient
choice is \cite{Pato} 
\begin{equation}
\Gamma _2 \left( x \right) \equiv \int\limits_{C_{\zeta x} } {dz^\nu } A_\nu
\left( z \right) \equiv \int\limits_0^1 {d\lambda x^i } A_i \left( {\lambda
x } \right) = 0,  \label{WLIV20}
\end{equation}
where $\lambda$ $(0\leq \lambda\leq1)$ is the parameter describing the
spacelike straight path $x^i = \zeta ^i + \lambda \left( {x - \zeta}
\right)^i $ , and $\zeta $ is a fixed point (reference point). There is no
essential loss of generality if we restrict our considerations to $\zeta
^i=0 $. In this case, the only nonvanishing equal-time Dirac bracket is 
\begin{equation}
\left\{ {A_i \left( x \right),\Pi ^j \left( y \right)} \right\}^ * =\delta{\
_i^j} \delta ^{\left( 3 \right)} \left( {x - y} \right) - \partial _i^x
\int\limits_0^1 {d\lambda x^j } \delta ^{\left( 3 \right)} \left( {\lambda x
- y} \right).  \label{WLIV25}
\end{equation}

We are now in a position to evaluate the interaction energy between
pointlike sources in the model under consideration, where a fermion is
localized at $\mathbf{y}\prime$ and an antifermion at $\mathbf{y}$. From our
above discussion, we see that $\left\langle H \right\rangle _\Phi$ reads 
\begin{eqnarray}
\left\langle H \right\rangle _\Phi &=& \left\langle \Phi \right|\int {d^3 x} 
\Biggl[\frac{1}{2}\Pi _i\left( {\frac{{\nabla ^4 - m_1 \nabla ^2 + m_2^2 }}{{\nabla ^4 - \xi ^2 \nabla ^2 + \rho ^2 }}} \right)\Pi ^i -\frac{Q}{2}\Pi _i \left( {\frac{{\nabla ^2 - m_2 }}{{\nabla ^4 - \xi ^2 \nabla ^2 + \rho ^2 }}} \right)B^i + QB_i \Pi ^i  \notag \\
&+& Q^2 B_i \frac{{\left( {\nabla ^2 - m_2 } \right)}}{{\left( {\nabla ^4 -
m_1 \nabla ^2 + m_2^2 } \right)}}B^i + \frac{1}{4}F_{ij}
\left( {\frac{{\nabla ^4 - a^2 \nabla ^2 + b^2 }}{{\nabla ^4 - m_1 \nabla ^2 +
m_2 ^2 }}} \right)F^{ij}\Biggr]|\Phi\rangle .  \label{WLIV30}
\end{eqnarray}

Next, as was first established by Dirac \cite{Dirac}, the physical state can
be written as 
\begin{equation}
\left| \Phi \right\rangle \equiv \left| {\overline \Psi \left( \mathbf{y }
\right)\Psi \left( \mathbf{y}\prime \right)} \right\rangle = \overline \psi
\left( \mathbf{y }\right)\exp \left( {iq\int\limits_{\mathbf{y}\prime}^{ 
\mathbf{y}} {dz^i } A_i \left( z \right)} \right)\psi \left(\mathbf{y}\prime
\right)\left| 0 \right\rangle,  \label{WLIV35}
\end{equation}
where $\left| 0 \right\rangle$ is the physical vacuum state and the line
integral appearing in the above expression is along a spacelike path
starting at $\mathbf{y}\prime$ and ending at $\mathbf{y}$, on a fixed time
slice. From this we see that the fermion fields are now dressed by a cloud
of gauge fields.

From the foregoing Hamiltonian structure we then easily verify that 
\begin{equation}
\Pi _i \left( x \right)\left| {\overline \Psi \left( \mathbf{y }\right)\Psi
\left( {\mathbf{y}^ \prime } \right)} \right\rangle = \overline \Psi \left( 
\mathbf{y }\right)\Psi \left( {\mathbf{y}^ \prime } \right)\Pi _i \left( x
\right)\left| 0 \right\rangle + q\int_ {\mathbf{y}}^{\mathbf{y}^ \prime } {\
dz_i \delta ^{\left( 3 \right)} \left( \mathbf{z - x} \right)} \left| \Phi
\right\rangle.  \label{WLIV40}
\end{equation}
Having made this observation and since the fermions are taken to be
infinitely massive (static) we can substitute $\Delta$ by $- \nabla ^2$ in
Eq. (\ref{WLIV30}). In such a case $\left\langle H \right\rangle _\Phi$
reduces to 
\begin{equation}
\left\langle H \right\rangle _\Phi = \left\langle H \right\rangle _0 +
\left\langle H \right\rangle _\Phi ^{\left( 1 \right)} + \left\langle H
\right\rangle _\Phi ^{\left( 2 \right)} ,  \label{WLIV45}
\end{equation}
where $\left\langle H \right\rangle _0 = \left\langle 0 \right|H\left| 0
\right\rangle$, and the $\left\langle H \right\rangle _\Phi ^{\left( 1
\right)}$ and $\left\langle H \right\rangle _\Phi ^{\left( 2 \right)}$ terms
are given by 
\begin{equation}
\left\langle H \right\rangle _\Phi ^{\left( 1 \right)} = - \frac{1} {{4
\sqrt {1 - {\raise0.7ex\hbox{${4\rho ^2 }$} \!\mathord{\left/ {\vphantom
{{4\rho ^2 } {\xi ^4 }}}\right.\kern-\nulldelimiterspace} \!\lower0.7ex
\hbox{${\xi ^4
}$}}} }}\left\langle \Phi \right|\int {d^3 x} \Pi _i \left[ {\left( {1 + 
\sqrt {1 - {\raise0.7ex\hbox{${4\rho ^2 }$} \!\mathord{\left/ {\vphantom
{{4\rho ^2 } {\xi ^4 }}}\right.\kern-\nulldelimiterspace} \!\lower0.7ex
\hbox{${\xi ^4 }$}}} } \right)\frac{{\nabla ^2 }} {{\left( {\nabla ^2 -
M_1^2 } \right)}} - \left( {1 - \sqrt {1 - {\raise0.7ex\hbox{${4\rho ^2 }$}
\!\mathord{\left/ {\vphantom {{4\rho ^2 } {\xi ^4
}}}\right.\kern-\nulldelimiterspace} \!\lower0.7ex\hbox{${\xi ^4 }$}}} }
\right)\frac{{\nabla ^2 }}{{\left( {\nabla ^2 - M_2^2 } \right)}}} \right]
\Pi ^i \left| \Phi \right\rangle,  \label{WLIV50}
\end{equation}

\begin{equation}
\left\langle H \right\rangle _\Phi ^{\left( 2 \right)} = \frac{1} {{2\left( {
M_1^2 - M_2^2 } \right)}}\left\langle \Phi \right|\int {d^3 x} \Pi _i \left[ 
{\left( {m_1 M_1^2 - m_2^2 } \right)\frac{1}{{\left( {\nabla ^2 - M_1^2 }
\right)}} + \left( {m_2^2 - m_1 M_2^2 } \right)\frac{1}{{\left( {\nabla ^2 -
M_2^2 } \right)}}} \right]\Pi ^i \left| \Phi \right\rangle,  \label{WLIV55}
\end{equation}
with $M_1^2 \equiv {\textstyle{\frac{1 }{2}}}\left( {\xi ^2 + \sqrt {\xi ^4
- 4\rho ^2 } } \right)$ and $M_2^2 \equiv {\textstyle{\frac{1 }{2}}}\left( {
\xi ^2 - \sqrt {\xi ^4 - 4\rho ^2 } } \right)$, $M_1\geq M_2$.

Using Eq.(\ref{WLIV40}), we see that the potential for two opposite charges
located at $\mathbf{y}$ and $\mathbf{y^{\prime }}$ takes the form 
\begin{eqnarray}
V &=& - \frac{{q^2 }}{{4\pi \sqrt {1 - {\raise0.5ex\hbox{$\scriptstyle
{4\rho ^2 }$} \kern-0.1em/\kern-0.15em \lower0.25ex
\hbox{$\scriptstyle {\xi ^4
}$}}} }}\left\{ {\frac{{\left( {1 + \sqrt {1-{\raise0.5ex\hbox{$\scriptstyle
{4\rho ^2 }$} \kern-0.1em/\kern-0.15em \lower0.25ex
\hbox{$\scriptstyle {\xi ^4
}$}}} } \right)}}{2}\frac{{e^{ - M_1 L} }}{L} - \frac{{\left( {1 - \sqrt {1
- {\raise0.5ex\hbox{$\scriptstyle {4\rho ^2
}$}\kern-0.1em/\kern-0.15em \lower0.25ex\hbox{$\scriptstyle {\xi ^4
}$}}}} \right)}}{2}\frac{{e^{ - M_2 L} }}{L}} \right\}  \notag \\
&+& \frac{{q^2 }}{{8\pi \sqrt {\xi ^4 - 4\rho ^2 } }}\left\{ {\left( {m_1
M_1^2 - m_2^2 } \right)\ln \left( {1 + \frac{{\Lambda ^2 }}{{M_1^2 }}}
\right) + \left( {m_2 - m_1 M_2^2 } \right)\ln \left( {1 + \frac{{\Lambda ^2 
}}{{M_2^2 }}} \right)} \right\}L,  \label{WLIV70}
\end{eqnarray}
where $\Lambda$ is a cutoff and $|\mathbf{y}-\mathbf{y}^{\prime }|\equiv L$.
Expression (\ref{WLIV70}) immediately shows that the effect of including
condensates is a linear potential, leading to the confinement of static
charges. It is also easy to see that the same result is obtained in the
timelike case. Before going ahead, we would like to remark how to give a
meaning to the would-be cutoff $\Lambda $. To do that, we should recall that
our effective model for the electromagnetic field is an effective
description that comes out upon integration over the $\Lambda $-field, whose
excitations are massive (Recall that $\Gamma=0$ for $p^2=M_1^2$ and $
p^2=M_2^2$). $1/M_1$ and $1/M_2$, the Compton wavelengths of these
excitations, naturally define a correlation distance. Physics at distances
of the order or lower than $1/M_2$ must necessarily take into account a
microscopic description of axion fields. This means that, if we work with
energies of the order or higher than $M_2$, our effective description with
the integrated effects of $\Lambda $ is no longer sensible. So, it is
legitimate that, for the sake of our analysis, we identify $\Lambda $ with $
M_1$. Then, with this identification, the potential of Eq. (\ref{WLIV70})
takes the form below: 
\begin{eqnarray}
V &=& - \frac{{q^2 }}{{4\pi \sqrt {1 - {\raise0.5ex\hbox{$\scriptstyle
{4\rho ^2 }$} \kern-0.1em/\kern-0.15em \lower0.25ex
\hbox{$\scriptstyle {\xi ^4
}$}}} }}\left\{ {\frac{{\left( {1 + \sqrt {1-{\raise0.5ex\hbox{$\scriptstyle
{4\rho ^2 }$} \kern-0.1em/\kern-0.15em \lower0.25ex
\hbox{$\scriptstyle {\xi ^4
}$}}} } \right)}}{2}\frac{{e^{ - M_1 L} }}{L} - \frac{{\left( {1 - \sqrt {1
- {\raise0.5ex\hbox{$\scriptstyle {4\rho ^2
}$}\kern-0.1em/\kern-0.15em \lower0.25ex\hbox{$\scriptstyle {\xi ^4
}$}}}} \right)}}{2}\frac{{e^{ - M_2 L} }}{L}} \right\}  \notag \\
&+& \frac{{q^2 }}{{8\pi \sqrt {\xi ^4 - 4\rho ^2 } }}\left\{ {\left( {m_1
M_1^2 - m_2^2 } \right)\ln \left( 2 \right) + \left( {m_2 - m_1 M_2^2 }
\right)\ln \left( {1 + \frac{{M_1^2 }}{{M_2^2 }}} \right)} \right\}L.
\label{WLIV75}
\end{eqnarray}
It is appropriate to observe the presence of a finite string tension in
Eq. (\ref{WLIV75}).

\section{Concluding remarks}

Our efforts in this contribution have focused on the possibility to realize a specific model of LSV in a scenario dominated by SUSY. We propose that the scale where LSV occurs is high enough to accommodate SUSY: either SUSY is exact when LSV takes place or the latter happens very close to the SUSY breaking scale. Our viewpoint is that exact SUSY effects or SUSY imprints may interfere with the breaking of relativistic invariance. We actually choose to assess LSV in a framework with exact SUSY. 

With this working hypothesis, we place the background responsible for LSV in a specific N=$1$-D= $4$-SUSY supermultiplet, namely, the chiral scalar superfield. It naturally extends the so-called Carroll-Field-Jackiw realization of LSV and brings about a neutral fermionic background field, described by a Majorana fermion. We are able to keep track of the effects of such a fermionic background through the condensates it may yield which severely affect the photino dispersion relations, as shown in eq. (\ref{III-75f}).

Having in mind the investigation of an effective model for photons induced by the effects of SUSY in our framework with LSV, we choose to integrate out the photino degrees of freedom and we end up with an effective Lagrangian for the photonic sector in which the non-trivial combinations of the fermion condensates explicit appear as coefficients of the terms in $FF$ and $F\tilde F$.

 This effective model is discussed in connection with the attainment of a static potential for the interaction between opposite electrically charged particles and its main features have been discussed which also account for the effects of the condensates induced by SUSY in the process of breaking Lorentz symmetry. 

 So, to conclude, we highlight that, according to our approach to the problem of probing LSV in a supersymmetric framework, SUSY is naturally broken whenever a single chiral scalar supermultiplet triggers LSV. The signal for the breaking of SUSY lies on the emergence of a Goldstone fermion that propagates as a perturbation around the background that condensates to set up the breaking of the relativistic invariance. Also, as a byproduct, the photon and photino masses are split, which confirms SUSY violation.

Our future steps, and this shall be the matter of a forthcoming contribution, consist on the discussion of the LSV in connection with the so-called vector supermultiplet (in simple and extended SUSY), which is characterized by a richer family of fermion fields and, upon comparison of the patterns of LSV in diverse SUSY-dominated scenarios, we expect to adopt the constraints already known on the Lorentz-symmetry violating parameters to also get information on the scale of SUSY breaking through the SUSY fermion condensates.

\section{acknowledgements}

The authors (HB, LDB and JAH-N) express their gratitude to the Conselho Nacional de Desenvolvimento Cient\'{\i}fico e Tecnol\'{o}gico (CNPq-Brazil) for the financial support. One of us (PG) wishes to thank the Field Theory Group of CBPF for hospitality.


\begin{thebibliography}{99}
\bibitem{Colladay} D. Colladay and V. A. Kosteleck\'{y}, \textit{Phys. Rev. }
D \textbf{55}, 6760 (1997); D. Colladay and V. A. Kosteleck\'{y}, \textit{\
Phys. Rev. }D \textbf{58}, 116002 (1998).

\bibitem{Samuel} V. A. Kostelecky and S. Samuel, \textit{Phys. Rev. Lett}. 
\textbf{63}, 224 (1989); \textit{Phys. Rev. Lett}. \textbf{66}, 1811 (1991); 
\textit{Phys. Rev. }D\textbf{\ 39}, 683 (1989); \textit{Phys. Rev. }D\textbf{
\ \ 40}, 1886 (1989), V. A. Kostelecky and R. Potting, \textit{Nucl. Phys.}
B \textbf{\ 359}, 545 (1991); Phys. Lett. B \textbf{381}, 89 (1996); V. A.
Kostelecky and R. Potting, \textit{Phys. Rev. }D\textbf{\ 51}, 3923 (1995).

\bibitem{photons1} C. Adam and F. R. Klinkhamer, \textit{Nucl. Phys. }B 
\textbf{607}, 247 (2001); \textit{Nucl. Phys. }B \textbf{657}, 214 (2003);
Phys. Lett. B\textbf{\ 513}, 245 (2001); C. Kaufhold and F. R. Klinkhamer, 
\textit{Nucl. Phys. }B \textbf{734}, 1 (2006); R. Montemayor and L.F.
Urrutia, Phys. Rev. D \textbf{72}, 045018 (2005); X. Xue and J. Wu, Eur.
Phys. J. C \textbf{48}, 257 (2006); A.A. Andrianov and R. Soldati, \textit{\
Phys. Rev.} D\textbf{\ 51}, 5961 (1995); \textit{Phys. Lett. }B\textbf{\ 435}
, 449 (1998); A.A. Andrianov, R. Soldati and L. Sorbo, \textit{Phys. Rev.} D 
\textbf{\ 59}, 025002 (1999); Q. G. Bailey and V. Alan Kostelecky, Phys.
Rev. D \textbf{70}, 076006 (2004); M. Frank and I. Turan, Phys. Rev. D 
\textbf{74}, 033016 (2006); ~B. Altschul, \textit{Phys. Rev.} D \textbf{72},
085003 (2005); R. Lehnert and R. Potting, \textit{Phys. Rev. Lett}. \textbf{
\ 93}, 110402 (2004); R. Lehnert and R. Potting, \textit{Phys. Rev. D} 
\textbf{\ 70}, 125010 (2004); ~B. Altschul, ~\textit{Phys. Rev. Lett.} 
\textbf{98}, 041603 (2007); Phys.Rev. D \textbf{72, } 085003 (2005);
Phys.Rev. D \textbf{\ 73}, 036005 (2006); Phys. Rev. D \textbf{73}, 045004
(2006).

\bibitem{photons2} Cantcheff M. B \textit{Eur. Phys. J. C }\textbf{46, }247 - 254, (2006), H. Belich, et al. \textit{REV. BRAS. ENS. FIS. } \textbf{29}, 57 (2007).

\bibitem{Radiative} R. Jackiw and V. A. Kosteleck\'{y}, \textit{Phys. Rev.
Lett.} \textbf{82}, 3572 (1999); J. M. Chung and B. K. Chung \textit{Phys.
Rev. }D \textbf{63}, 105015 (2001); J.M. Chung, \textit{Phys.Rev}. D\textbf{
\ \ 60}, 127901 (1999); G. Bonneau, \textit{Nucl.Phys.} B\textbf{\ 593, }398
(2001); M. Perez-Victoria, \textit{Phys. Rev. Lett}. \textbf{83}, 2518
(1999); M. Perez-Victoria,\textit{\ JHEP} \textbf{0104}, 032 (2001); O.A.
Battistel and G. Dallabona, \textit{Nucl. Phys. B} \textbf{610}, 316 (2001);
O.A. Battistel and G. Dallabona, \textit{J. Phys. G} \textbf{28}, L23
(2002); \textit{J. Phys. G} \textbf{27}, L53 (2002); A. P. Ba\^{e}ta
Scarpelli, M. Sampaio, M. C. Nemes, and B. Hiller, \textit{Phys. Rev.} D 
\textbf{\ 64}, 046013 (2001); F.A. Brito, T. Mariz, J.R. Nascimento, E.
Passos, R.F. Ribeiro, JHEP 0510 (2005) 019.

\bibitem{fermions} B. Altschul, \textit{Phys. Rev. }D \textbf{70}, 056005
(2004); G. M. Shore, Nucl. Phys. \textbf{B} 717, 86 (2005); \ D. Colladay
and V. A. Kostelecky, \textit{Phys. Lett. }B \textbf{511}, 209 (2001); M. M.
Ferreira Jr, P\textit{hys. Rev.} D \textbf{70,} 045013 (2004); M. M.
Ferreira Jr, P\textit{hys. Rev.} D \textbf{71,} 045003 (2005); M. M.
Ferreira Jr and M. S. Tavares, Int. J. Mod. Phys. \textbf{A} 22, \ 1685
(2007);J. Alfaro, A.A. Andrianov, M. Cambiaso, P. Giacconi, R. Soldati,
Phys. Lett. \textbf{B} 639 (2006) 586-590.

\bibitem{neutrinos} V. Barger, S. Pakvasa, T.J.Weiler, and K.Whisnant, 
\textit{Phys. Rev}. \textit{Lett.}\textbf{\ 85}, 5055 (2000); V. A.
Kostelecky and M. Mewes, \textit{Phys. Rev}. D\textbf{\ 69}, 016005 (2004); 
\textit{Phys. Rev}. D\textbf{\ 70}, 031902 (R) (2004); \textit{Phys. Rev}. D 
\textbf{\ 70}, 076002 (2004); T. Katori, A. Kostelecky, R. Tayloe, Phys.Rev.
D \textbf{74}, 105009 (2006); R. Brustein, D. Eichler, S. Foffa, Phys.Rev. D 
\textbf{65}, 105006 (2002); Y. Grossman,C. Kilic, J. Thaler, and D.G. E.
Walker, Phys. Rev. D \textbf{72}, 125001 (2005); L. B. Auerbach \textit{et
al.}, Phys. Rev. D \textbf{72}, 076004 (2005); D. Hooper, D. Morgan, and E.
Winstanley, Phys. Rev. D \textbf{72}, 065009 (2005).

\bibitem{Defects} M. Lubo, \textit{Phys. Rev.} D \textbf{71}, 047701 (2005);
M.N. Barreto, D. Bazeia, and R. Meneses, \textit{Phys. Rev. D} \textbf{73},
065015 (2006).

\bibitem{Phases} H. Belich, T. Costa-Soares, M.M. Ferreira Jr., J. A. Helay 
\"{e}l-Neto, M.T. D. Orlando, \textit{Phys. Lett. B }\textbf{639, }678
(2006); H. Belich, E. O. Silva, M. M. Ferreira Jr. and M. T. D. Orlando,
Phys. Rev. D 83, 125025 (2011); K. Bakke and H. Belich, Eur.
Phys. J. Plus 127, 102 (2012);  K. Bakke, H. Belich and E. O. Silva,
Ann. Phys. (Berlin) 523, 910 (2011); K. Bakke, et al., J. Math. Phys. 52, 063505 (2011); K. Bakke and H. Belich, J. Phys. G: Nucl. Part. Phys. 39, 085001 (2012);  K. Bakke, E. O. Silva and H. Belich, J. Phys. G: Nucl. Part. Phys. 39, 055004 (2012); H. Belich, et al., Phys. Rev. D 86, 125037 (2012).
 
\bibitem{CosmicRay} O. Gagnon and G. D. Moore, \textit{Phys. Rev. D} \textbf{
\ 70}, 065002 (2004); J.W. Moffat, \textit{Int. J. Mod. Phys. D }\textbf{12}
1279 (2003); F. W. Stecker and S.T. Scully, Astropart. Phys. \textbf{23},
203 (2005); F. W. Stecker and S.L. Glashow, Astropart. Phys. \textbf{16}, 97
(2001).

\bibitem{Iltan} E. O. Iltan, Eur. Phys. J. C \textbf{40}, 269 (2005); Mod.
Phys. Lett. A19, 327 (2004); JHEP 0306 (2003) 016.

\bibitem{Lehnert1} V.A. Kostelecky and R. Lehnert, Phys. Rev. D\textbf{\ 63,}
065008 (2001); R. Lehnert, \textit{Phys. Rev.} D \textbf{68}, 085003 (2003).

\bibitem{General} H. Belich, M. A. De Andrade, M. A. Santos, Mod.Phys.Lett.
A \textbf{20}, 2305 (2005)H. Belich, T. Costa-Soares, M.M. Ferreira Jr., J.
A. Helay\"{e}l-Neto and M.T.D. Orlando, Int. J. Mod. Phys. \textbf{A 21},
2415 (2006), A. P. Baeta Scarpelli and J. A. Helayel-Neto, Phys. Rev. D 
\textbf{73}, 105020 (2006); N.M. Barraz, Jr., J.M. Fonseca, W.A. Moura-Melo,
and J.A. Helayel-Neto, Phys. Rev. D\textbf{76}, 027701 (2007),H.\ Belich 
\textit{et al}., \textit{Phys. Lett}. A\textbf{\ 370,} 126 - 130 (2007), R.
Casana, M. M. Ferreira,J. S. Rodrigues, \textit{Phys. Rev}. D\textbf{\ 78},
125013-1 (2008), R. Casana \textit{et al}.,\textit{Phys. Rev}. D\textbf{\ 78}
, 105014 (2008) , R. Casana \textit{et al}.,\textit{Phys. Rev}. D\textbf{\
78 }, 025030(2008), R. Casana \textit{et al}.,\textit{Phys. Rev}. D\textbf{\
80} , 125040(2009), R. Casana \textit{et al}., \textit{Phys. Rev}. D\textbf{
\ 80} , 085026(2009), R. Casana \textit{et al}., \textit{Eur. Phys. J. C }
\textbf{\ 62, }573-578, (2009).

\bibitem{Tests} S.R. Coleman and S.L. Glashow, \textit{Phys. Rev}. D\textbf{
\ 59}, 116008 (1999); V. A. Kostelecky and M. Mewes, \textit{Phys. Rev. Lett}
. \textbf{87}, 251304 (2001); V. A. Kostelecky and M. Mewes, \textit{Phys.
Rev}. D\textbf{\ 66}, 056005 (2002); J. Lipa et. al., \textit{Phys. Rev.
Lett.} \textbf{90}, 060403 (2003).

\bibitem{CPT} R. Bluhm, V.A. Kostelecky, and N. Russell, \textit{Phys. Rev.
Lett}. \textbf{79}, 1432 (1997); R. Bluhm, V.A. Kostelecky, and N. Russell, 
\textit{Phys. Rev. }D \textbf{57}, 3932 (1998); R. Bluhm, V.A. Kostelecky,
C. D. Lane, and N. Russell, \textit{Phys. Rev. Lett}. \textbf{88}, 090801
(2002); R. Bluhm and V.A. Kostelecky, \textit{Phys. Rev. Lett}. \textbf{84},
1381 (2000); R. Bluhm, R. Bluhm, V.A. Kostelecky, and C. D. Lane, \textit{
Phys. Rev. Lett}. \textbf{84}, 1098 (2000); V.A. Kostelecky, and N. Russell, 
\textit{Phys. Rev. Lett}. \textbf{82}, 2254 (1999); V.A. Kostelecky and C.
Lane, \textit{Phys. Rev. }D \textbf{60}, 116010 (1999).

\bibitem{Hamilton} V.A. Kostelecky and C. D. Lane, J. Math. Phys. 40, 6245
(1999); R. Lehnert, \textit{J. Math. Phys.} \textbf{45}, 3399 (2004).

\bibitem{Manojr} M.M.\ Ferreira Jr and F. M. O. Moucherek, Int. J. Mod.
Phys. \textbf{A} 21, 6211\ (2006); Nucl. Phys. A \textbf{790}, 635 (2007);
S. Chen, B. Wang, and R. Su, Class. Quant.Grav. \textbf{23}, 7581,(2006);O.
G. Kharlanov and V. Ch. Zhukovsky, \textit{J. Math. Phys.} \textbf{48},
092302 (2007), M. Frank, I. Turan, I. Yurdusen, \textit{hep-th} 07094276.

\bibitem{Cavity} H. Muller, C. Braxmaier, S. Herrmann, and A. Peters, 
\textit{Phys. Rev.} D\textbf{\ 67}, 056006 (2003); H. M\"{u}ller, A. Saenz,
A. Peters, and C. L\"{a}mmerzahl, P\textit{hys. Rev.} D \textbf{68, }116006
(2003); H. M\"{u}ller, \textit{Phys. Rev.} D\textbf{\ 71}, 045004 (2005); N.
Russell, Phys. Scripta \textbf{72}, C38 (2005).

\bibitem{Masers} D. F. Phillips, M. A. Humphrey, E. M. Mattison, R. E.
Stoner, R. F C. Vessot, R. L. Walsworth, Phys.Rev. D63, 111101 (2001); D.
Bear, R.E. Stoner, R.L. Walsworth, V. Alan Kostelecky, Charles D. Lane,
Phys.Rev. Lett. \textbf{85}, 5038 (2000); Erratum-ibid. 89 (2002) 209902;
M.A. Humphrey, D.F. Phillips, R.L. Walsworth, Phys. Rev. A 68, 063807 (2003).

\bibitem{LSVSUSY} D. Colladay  and  P. MacDonald, Phys.Rev. D83 (2011) 025021;  M. S. Berger and V. A. Kostelecky,
Phys. Rev. D 65, 091701(2002); C.F. Farias,  U.,  A.C. Lehum,  J.R. Nascimento and A.Yu. Petrov, Phys.Rev. D86 (2012) 065035;  P.  A. Bolokhov, S. Nibbelink Groot  and  M. Pospelov, Phys.Rev. D72 (2005) 015013;  A.  Katz and Y. Shadmi  Phys.Rev. D74 (2006) 115021;  S. Groot Nibbelink  and  M. Pospelov,  Phys.Rev.Lett. 94 (2005) 081601; "A Discussion on the role of fermionic supersymmetric partners in Lorentz-symmetry breaking", J. A. Helay\"el-Neto, H. Belich, G.S. Dias, F.J.L. Leal and  W. Spalenza, Proceedings of Science, PoS - ICFI2010 (2010) 032. 
 
\bibitem{David} D. Redigolo, Phys.Rev. D85 (2012) 085009.

 \bibitem{GaeteEuro3} P. Gaete and E. Spallucci, Phys. Lett. B \textbf{675}, 145 (2009).

\bibitem{GaeteHel} P. Gaete and J. A. Helayel-Neto, Phys. Lett. B
 \textbf{683}, 211 (2010).
 
 \bibitem{CFJ} S. Carroll, G. Field and R. Jackiw, Phys. Rev. D
 \textbf{41}, 1231 (1990). 
 
\bibitem{GaeteEuro1} P. Gaete and E. Spallucci, Phys. Rev. D \textbf{77}, 027702 (2008).

\bibitem{GaeteEuro2} P. Gaete and E. Spallucci, J. Phys. A: Math. Theor. \textbf{41}, 185401 (2008).

\bibitem{Baeta}  A.P. Baeta Scarpelli, H. Belich, J.L. Boldo, L.P. Colatto, J.A. Helay\"el-Neto, A.L.M.A. Nogueira, Nucl.Phys.Proc.Suppl. 127 (2004) 105-109.

\bibitem{GaeteSch} P. Gaete and I. Schmidt, Phys. Rev. D \textbf{61}, 125002 (2000).

\bibitem{Gaete99} P. Gaete, Phys. Rev. D \textbf{59}, 127702 (1999).

\bibitem{Pato} P. Gaete, Z. Phys. C \textbf{76}, 355 (1997).

\bibitem{Dirac} P. Dirac, Can. J. Phys. \textbf{33}, 650 (1955).
\end{thebibliography}
\end{document}